\documentclass[twocolumn,superscriptaddress,prl,aps,noeprint]{revtex4-1}
\usepackage{graphicx}
\usepackage{times}
\usepackage{textcomp}
\usepackage{amsmath}
\usepackage{amssymb}
\usepackage{units}
\usepackage{stmaryrd} 
\usepackage{braket}
\usepackage{bm}
\usepackage[compact]{titlesec}
\titlespacing{\section}{0pt}{*0}{*0}
\titlespacing{\subsection}{0pt}{*0}{*0}
\titlespacing{\subsubsection}{0pt}{*0}{*0}

\renewcommand{\vec}[1]{\boldsymbol{#1}}

\begin{document}
\preprint{0}

\title{Spin-Orbit-Controlled Metal-Insulator Transition in Sr$_2$IrO$_4$}

\author{B. Zwartsenberg}
\affiliation{Quantum Matter Institute, University of British Columbia, Vancouver, BC V6T 1Z4, Canada}
\affiliation{Department of Physics and Astronomy, University of British Columbia, Vancouver, BC V6T 1Z1, Canada}
\author{R.P. Day}
\affiliation{Quantum Matter Institute, University of British Columbia, Vancouver, BC V6T 1Z4, Canada}
\affiliation{Department of Physics and Astronomy, University of British Columbia, Vancouver, BC V6T 1Z1, Canada}
\author{E. Razzoli}
\affiliation{Quantum Matter Institute, University of British Columbia, Vancouver, BC V6T 1Z4, Canada}
\affiliation{Department of Physics and Astronomy, University of British Columbia, Vancouver, BC V6T 1Z1, Canada}
\author{M. Michiardi}
\affiliation{Quantum Matter Institute, University of British Columbia, Vancouver, BC V6T 1Z4, Canada}
\affiliation{Department of Physics and Astronomy, University of British Columbia, Vancouver, BC V6T 1Z1, Canada}
\affiliation{Max Planck Institute for Chemical Physics of Solids, N{\"o}thnitzer Stra{\ss}e 40, 01187 Dresden, Germany}
\author{N. Xu}
\affiliation{Swiss Light Source, Paul Scherrer Institut, CH-5232 Villigen PSI, Switzerland}
\author{M. Shi}
\affiliation{Swiss Light Source, Paul Scherrer Institut, CH-5232 Villigen PSI, Switzerland}
\author{J.D. Denlinger}
\affiliation{Advanced Light Source, Lawrence Berkeley National Laboratory, Berkeley, California 94720, USA}
\author{G. Cao}
\affiliation{Department of Physics, The Ohio State University, Columbus, Ohio 43210, United States}
\author{S. Calder}
\affiliation{Neutron Scattering Division, Oak Ridge National Laboratory, Oak Ridge, Tennessee 37831, USA}
\author{K. Ueda}
\author{J. Bertinshaw}
\author{H. Takagi}
\affiliation{Max Planck Institute for Solid State Research, Heisenbergstra{\ss}e 1, 70569 Stuttgart, Germany}
\author{B.J. Kim}
\affiliation{Department of Physics, Pohang University of Science and Technology, Pohang 790-784, South Korea}
\affiliation{Max Planck Institute for Solid State Research, Heisenbergstra{\ss}e 1, 70569 Stuttgart, Germany}
\affiliation{Center for Artificial Low Dimensional Electronic Systems, Institute for Basic Science (IBS), 77 Cheongam-Ro, Pohang, 790-784, Republic of Korea}
\author{I.S. Elfimov}
\affiliation{Quantum Matter Institute, University of British Columbia, Vancouver, BC V6T 1Z4, Canada}
\affiliation{Department of Physics and Astronomy, University of British Columbia, Vancouver, BC V6T 1Z1, Canada}
\author{A. Damascelli}
\email{damascelli@physics.ubc.ca}
\affiliation{Quantum Matter Institute, University of British Columbia, Vancouver, BC V6T 1Z4, Canada}
\affiliation{Department of Physics and Astronomy, University of British Columbia, Vancouver, BC V6T 1Z1, Canada}

\date{\today}

\maketitle

{\bf In the context of correlated insulators, where electron-electron interactions (U) drive the localization of charge carriers, the metal-insulator transition (MIT) is described as either bandwidth (BC) or filling (FC) controlled \cite{Imada1998}. Motivated by the challenge of the insulating phase in Sr$_2$IrO$_4$, a new class of correlated insulators has been proposed, in which spin-orbit coupling (SOC) is believed to renormalize the bandwidth of the half-filled $j_{\mathrm{eff}} = 1/2$ doublet, allowing a modest U to induce a charge-localized phase \cite{Kim2008,Kim2009}. Although this framework has been tacitly assumed, a thorough characterization of the ground state has been elusive \cite{MorettiSala2014, Kim2017}. Furthermore, direct evidence for the role of SOC in stabilizing the insulating state has not been established, since previous attempts at revealing the role of SOC \cite{Qi2012,Lee2012} have been hindered by concurrently occurring changes to the filling \cite{Brouet2015,Cao2016,Louat2018}. We overcome this challenge by employing multiple substituents that introduce well defined changes to the signatures of SOC and carrier concentration in the electronic structure, as well as a new methodology that allows us to monitor SOC directly. Specifically, we study Sr$_2$Ir$_{1-x}$T$_x$O$_4$ (T = Ru, Rh) by angle-resolved photoemission spectroscopy (ARPES), combined with \emph{ab-initio} and supercell tight-binding calculations. This allows us to distinguish relativistic and filling effects, thereby establishing conclusively the central role of SOC in stabilizing the insulating state of Sr$_2$IrO$_4$. Most importantly, we estimate the critical value for spin-orbit coupling in this system to be $\lambda_c = 0.42 \pm 0.01$ eV, and provide the first demonstration of a spin-orbit-controlled MIT.}

The familiar tools of chemical doping and pressure have provided straightforward access to both FC and BC MIT in conventional correlated insulators. In an effort to unveil the role of SOC in the insulating behavior of Sr$_2$IrO$_4$, and whether it can indeed drive a MIT, we have attempted to controllably dilute SOC in the valence electronic structure by substituting Ir ($\lambda_{SOC} \sim$ 0.4 eV \cite{Mattheiss1976,Moon2008,Kim2012}) with Ru and Rh ($\lambda_{SOC} \sim$ 0.19 eV \cite{Haverkort2008, Veenstra2014,Earnshaw1961}). While these substituents have similar values of $\lambda_{SOC}$ and are both $4d$ ions with comparable values for $U$ \cite{Mravlje2011, Martins2011} and ionic radii \cite{Shannon1976}, they are otherwise distinct: Ru has one less electron than Rh, and is therefore associated with a markedly larger impurity potential. We will show through supercell tight-binding model calculations that this leads to a pronounced contrast in the consequences of Rh and Ru substitution: the larger impurity potential associated with Ru precludes a significant reduction of the valence SOC. By comparison, Rh is electronically more compatible with Ir, facilitating a successful dilution of SOC. We measure this evolution directly, through orbital mixing imbued by SOC, manifest experimentally in the photoemission dipole matrix elements. To comprehend all aspects of the MIT observed here for both Rh and Ru substitution, we consider individually the effects of filling (Fig. 1), correlations/bandwidth (Fig. 2), and spin-orbit coupling (Figs. 3 and 4), ultimately concluding that the transition in Sr$_2$Ir$_{1-x}$T$_x$O$_4$ is a spin-orbit controlled MIT.

\begin{figure}
\centering 
  \includegraphics[width = 3.4in]{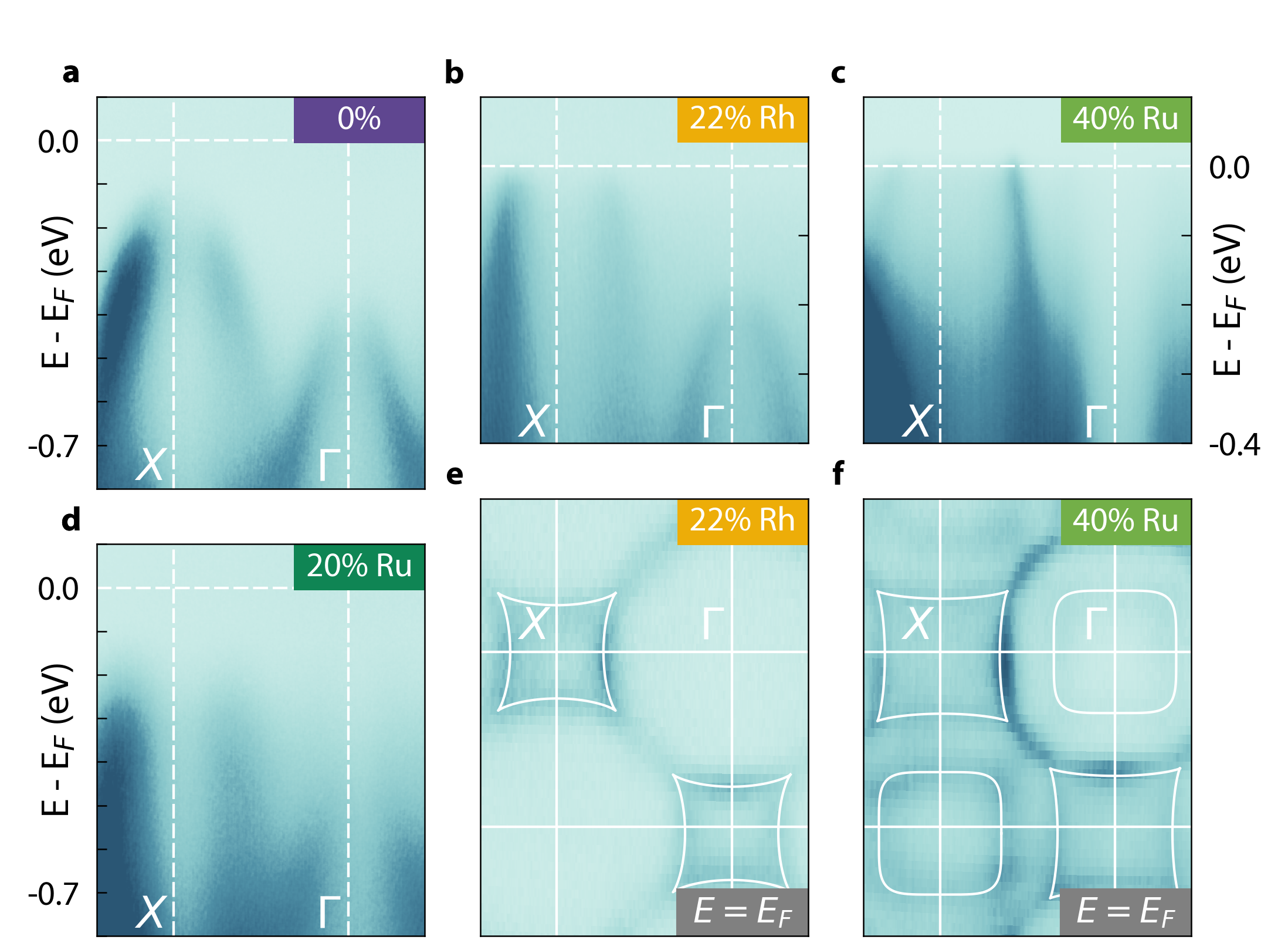}
 \caption{{\bf Dependence of the MIT on Rh and Ru substitution.} {\bf a-d} ARPES spectra along $\Gamma - X$ for the pristine sample, $x_{Rh} = 0.22$, $x_{Ru} = 0.40$ and $x_{Ru} = 0.20$, respectively. {\bf e} and {\bf f} show Fermi surface maps for $x = 0.22$ Rh and $x = 0.40$ Ru. The sizes of the pockets are indicated with white lines. Fermi surface maps are integrated over 50 meV. All data taken at $h\nu = 64$ eV with temperatures between 120 K and 150 K for $x \leq 0.10$, and below 40 K otherwise.}
\label{fig1}
\end{figure}

Having highlighted the three relevant aspects of the MIT, we begin our disquisition by showcasing the changes both substituents introduce to the electronic structure of Sr$_2$IrO$_4$ as measured by ARPES. Fig. 1a-d summarize ARPES spectra for $x = 0$, $x_{Rh} = 0.22$, and $x_{Ru} = 0.20,0.40$. As reported previously \cite{Kim2008}, the pristine sample supports an energy gap, with a band maximum at $X$ at a binding energy of around $E_b = 0.25$ eV. When substituting Rh, a pseudo-gapped metallic state forms for concentrations $x \gtrsim 0.13$ \cite{Brouet2015,Cao2016,Louat2018}. This is exemplified by our $x_{Rh} = 0.22$ data, shown in Fig. 1b,e. At comparable values of $x_{Ru}$, the system remains insulating (cf. $x_{Ru} = 0.20$ in Fig. 1d), and only by going as high as $x_{Ru} = 0.40$ (Fig. 1c,f) do we find that the MIT has been traversed \cite{Cava1994, Yuan2015, Wang2018}, consistent with transport measurements \cite{Yuan2015}. 

 Within the metallic phase, the Fermi surface volume provides a direct measure of the hole doping introduced by the impurity atoms. We report a Brillouin zone coverage of 16\% and 46\% for Rh and Ru respectively, corresponding to a nominal doping of 0.16 holes  (at $x_{Rh}$ = 0.22) and 0.46 holes (at $x_{Ru} = 0.40$), per formula unit. To within our level of certainty, each impurity atom then contributes approximately one hole carrier, with Ru perhaps contributing a somewhat larger number than Rh. This observation runs contrary to the expectations for a FC transition: despite contributing at least as many holes as Rh, the MIT critical concentration required for Ru is roughly double that of Rh. This precludes a transition described in terms of filling, despite earlier reports to the contrary \cite{Brouet2015,Cao2016,Louat2018}. An explanation in terms of the modification to the crystal structure upon Ru substitution can be equally excluded: the smaller ionic radius of Ru causes a minimal reduction of octahedral distortions ($12^{\circ}$ to $10^{\circ}$) up to the concentrations used in our study \cite{Yuan2015}. More importantly, such a reduction of distortions would increase the bandwidth \cite{Martins2010}, and the expected trend would be opposite to our observations. Alternatively, due to the presence of a sizable impurity potential for Ru (as discussed below), disorder effects could also be considered; however, recent studies regarding disorder in Mott systems point out that also such effects would push the critical concentration to lower values \cite{Wang2018, Heidarian2004}, precipitating once again an earlier onset of metallicity in the Ru-substituted compounds.

\begin{figure}
\centering 
  \includegraphics[width = 3.4in]{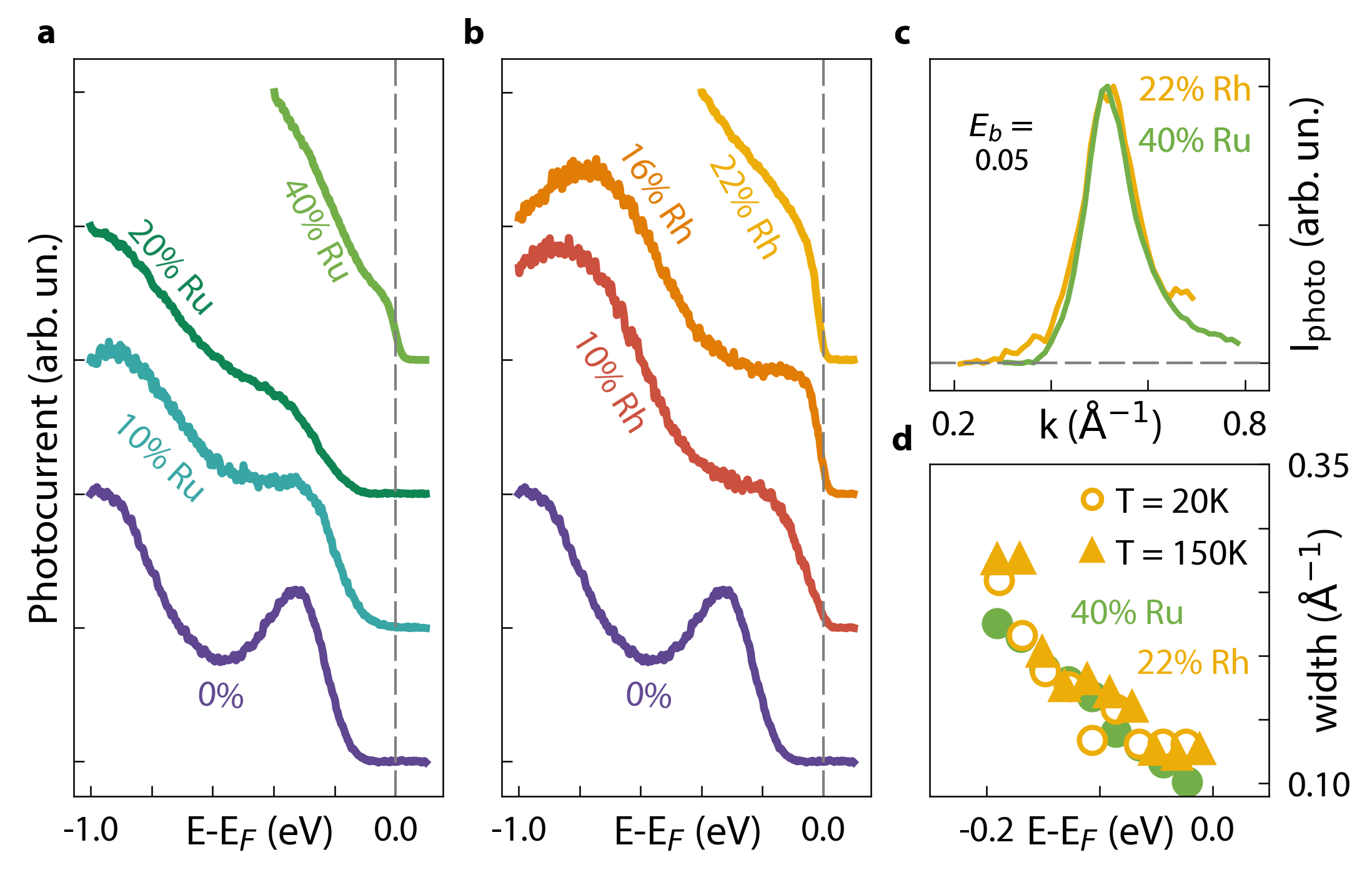}
 \caption{{\bf ARPES linewidth evolution with substitution.} Energy distribution curves (EDC's) for Ru {\bf a} and Rh {\bf b} substituted samples, taken at the momentum with the leading edge closest to the Fermi energy. Photon energies and temperatures for the EDCs are the same as in Fig. 1 {\bf c} Momentum distribution curve (MDC) curves for $x_{\mathrm{Ru}} = 0.40$ and $x_{\mathrm{Rh}} = 0.22$. {\bf d} MDC fits for $x_{\mathrm{Ru}} = 0.40$ and $x_{\mathrm{Rh}} = 0.22$. MDC data shown in {\bf c,d} were taken using $h\nu = 92$ eV at a temperature of 20 K.}
\label{fig2}
\end{figure}

Looking beyond the disparate critical concentrations associated with Ru and Rh substitution, analysis of the ARPES spectral features allows for a more thorough comparison of these materials to be made. The selected energy distribution curves (EDCs) cut through the valence band maximum for each doping in Fig. 2a (Ru) and 2b (Rh), reflecting the evolution of each material across the MIT. This coincides with a definitive Fermi level crossing in the EDCs of Fig. 2a and b, from which we can infer the critical concentrations to be $x_{\mathrm{Rh}} = 0.13 \pm 0.03$ and $x_{\mathrm{Ru}} = 0.30 \pm 0.10$ (this matches previous photoemission work on the Rh substituted compound \cite{Brouet2015,Cao2016,Louat2018}; as for the Ru-substituted samples, those have not previously been studied by photoemission). As the interpretation of EDC lineshape is non-trivial \cite{Kaminski2001}, we turn to an analysis of momentum distribution curves (MDCs) for a more quantitative analysis of the evolution of correlation effects. The MDC linewidth is directly related to the state lifetime, and by extension to both electronic interactions and disorder \cite{Damascelli2004, Hufner1995, Mahan1978}. Two representative MDCs are shown in Fig. 2c for $x_{\mathrm{Rh}} = 0.22$ and $x_{\mathrm{Ru}} = 0.40$. Widths from these, and other MDCs along the dispersion, are summarized in Fig. 2d. As can be inferred by the comparison of data from 20 K and 150 K, correlations -- rather than thermal broadening -- are the limiting factor in determining the MDC linewidth.  Consideration of both $x_{\mathrm{Ru}} = 0.40$ and $x_{\mathrm{Rh}} = 0.22$ reveal remarkably similar interaction effects in the two compounds, despite their significant differences in composition and disorder. In addition, while spectral broadening at high binding energies precludes a precise evaluation of the bandwidth, we estimate the latter to be constant to within 10\% over the range of Rh/Ru concentrations considered. 

We have thus determined that while doping effects are comparable for Ru and Rh, similar correlated metallic phases are observed at very different concentrations. To rectify this apparent contradiction, one must consider the context of the present MIT: it has been proposed that the correlated insulating phase in Sr$_2$IrO$_4$ is stabilized by the strong spin-orbit coupling. This motivates consideration of the role SOC plays in the MIT for both Ru- and Rh- substituted compounds. The low-energy influence of SOC can be characterized by an effective value in the valence band, determined by the hybridization between atomic species as demonstrated in Ref. \onlinecite{Weeks2011,Hu2012}. This effect could cause a reduction of SOC effects in the valence band as a function of (Ru,Rh) substitution. We find the reduction of SOC to be strongly dependent on the presence of an impurity potential, which limits hybridization of host and impurity states, ultimately curtailing the dilution of SOC effects (see Supplementary Information). In light of the reported electronic phase separation for the Ru compound \cite{Carter1995, Glamazda2014,Calder2016}, this suggests that such dilution of SOC may be more effective for Rh, providing a natural explanation for their disparate critical concentration in substituted Sr$_2$IrO$_4$ compounds. 

The model presented in the Supplementary Information to illustrate the mechanism of spin-orbit mixing, can be made quantitative for the Ru/Rh iridates through consideration of impurity-substituted supercell models. Using density-functional theory (DFT), at $x = 0.25$ substitution, in Fig. 3a we observe good overlap between the Rh and Ir projected density of states (DOS). This can be compared against the same scenario for Ru in Fig. 3b, where the substituent DOS is found to align poorly with Ir. Such an offset, observed most clearly through consideration of the centre of mass of the Ru-projected DOS, has been reported previously for similar substitutions \cite{Wadati2010,Levy2012}. Calculating the band's centre of mass in terms of the projected densities of states for both, we find an impurity potential for Ru of 0.3 eV, which is close to the number found in \cite{Wadati2010,Levy2012} (0.25 eV), and agrees with Wannier calculations (0.2 eV) performed on the same supercells. This establishes a reasonable starting point from which we can explore the influence of doping on SOC effects in more detail. This is carried out through development of a supercell tight-binding (TB) model. We expand a single iridium TB Hamiltonian (see Supplementary Information) to a 64 site supercell, randomly substituting a fraction $x$ of sites with an impurity atom. For the sake of simplicity, the impurities are assumed to differ from Ir in only their $\lambda_{SOC}$ (0.19 eV for both Ru and Rh, 0.45 for Ir), and onsite potential (0.0 eV for Rh and Ir, $0.25 \pm 0.05$ eV for Ru). Similarly, octahedral distortions and electron correlations are neglected to better illustrate the energy shift of the $j_\mathrm{eff}$ states. We have used the unfolding method \cite{Boykin2005, Ku2010, Haverkort2011, Popescu2012} to project bands into the original Brillouin zone. By averaging the resulting spectral function over 200 random configurations, we observe a smooth evolution of effective SOC in this system, which depends strongly on the impurity potential.

\begin{figure}
\centering 
  \includegraphics[width = 3.4in]{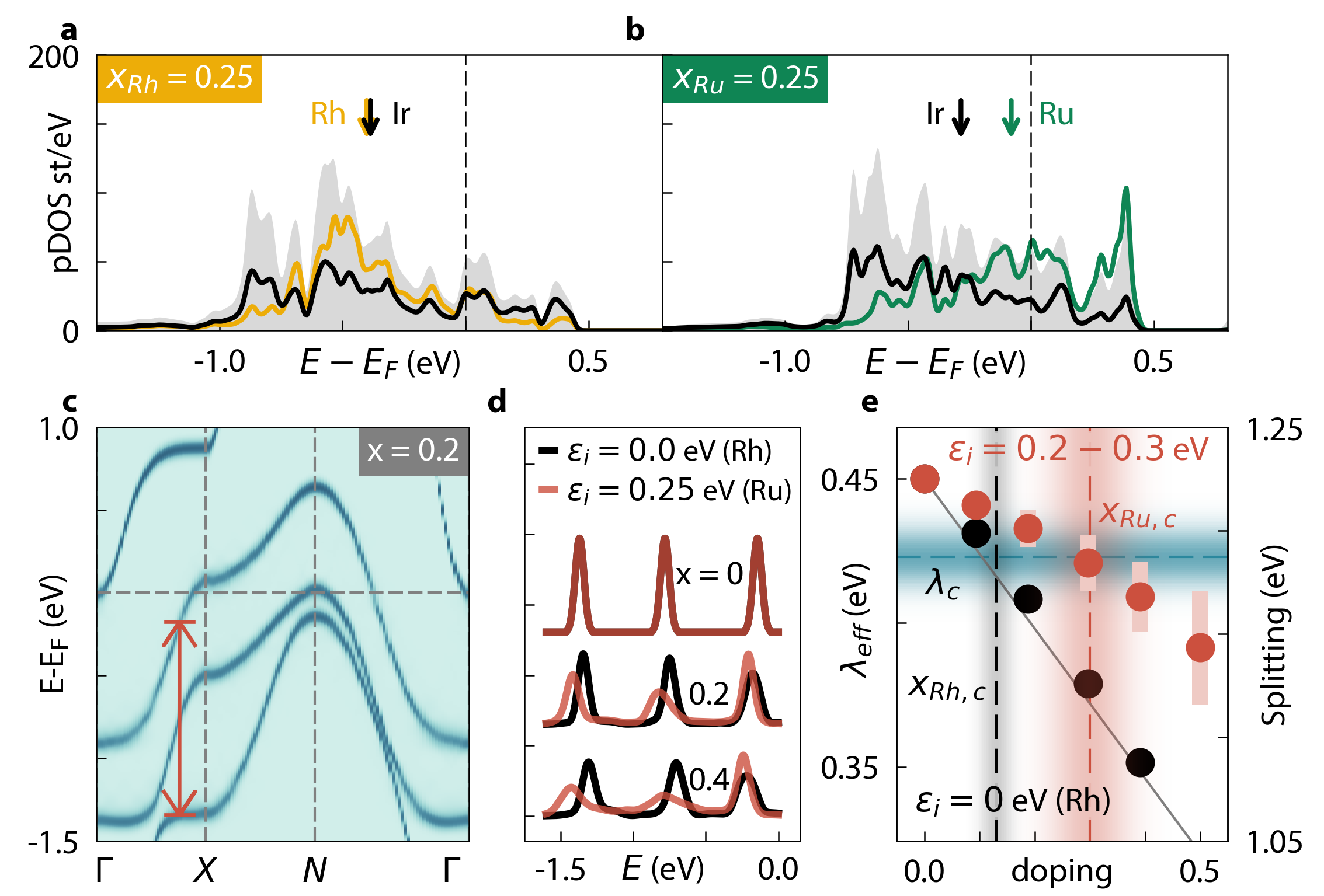}
 \caption{{\bf Reduction of SOC through supercell analysis.}  In {\bf a} and {\bf b}, an analysis of the impurity potential of Rh and Ru in Sr$_2$IrO$_4$ is plotted, as calculated by density-functional theory. The grey background represents the total DOS, normalized by the number of TM sites. The black curves show the Ir projected DOS per Ir ion in the 25\% substituted calculation, while the orange and green colored curves reflect the projected DOS per substituent ion for Rh and Ru respectively. The arrows indicate the center of mass for the projected bands. {\bf c} Supercell calculated spectrum for $x_{Rh} = 0.2$ obtained after unfolding. The effective splitting is indicated in red. {\bf d} Cuts (EDCs) for different concentrations of dopants at the position of the red arrow in {\bf c} [${\bf k}  = \left(\frac{3\pi}{4a}, 0\right)$]. The substituted atoms have different SOC  and on-site energy. We use $\lambda_{SOC} = 0.19$ eV and $\varepsilon_i = 0.0$ eV for Rh (black), while we use $\lambda_{SOC} = 0.19$ eV and $\varepsilon_i = 0.25$ eV for Ru (red). {\bf e} Progression of the splitting between the outermost peaks for simulations in {\bf d} for Rh (black markers) and Ru (red markers). For Rh, a linear interpolation is plotted between the end members of the phase diagram. For Ru, the resulting range of splitting for $\varepsilon_i = 0.25 \pm 0.05$ is indicated by red shaded rectangles. The critical concentrations obtained from our measurements are indicated by the black (Rh) and red (Ru) vertical shaded areas. The blue shaded area indicates the inferred $\lambda_{c} = 0.42 \pm 0.01$.}
\label{fig3}
\end{figure}

The results are summarized in Fig. 3, with a representative unfolded spectrum ($x_{Rh} = 0.20$) plotted in Fig. 3c. We investigate the level spacing at ${\bf k}  = \left(\frac{3\pi}{4a}, 0\right)$, indicated by the vertical arrow. This is the $k$-point at which we will later present experimental data. The change in splitting is seen clearly in Fig. 3d, where we present a series of EDCs at ${\bf k}  = \left(\frac{3\pi}{4a}, 0\right)$, for models with a non-zero on-site impurity potential (Ru, red), and those without (Rh, black). This doping dependence is summarized in Fig. 3e. The right vertical axis reflects the splitting observed at ${\bf k} = \left(\frac{3\pi}{4a}, 0\right)$, and the left the value of $\lambda_{SOC}$ that would produce the corresponding splitting in a model without substitutions (i.e. for an overall uniform value of $\lambda_{SOC}$). This second axis serves to illustrate the \emph{effective} spin-orbit coupling caused by substitution of Ir with Rh and Ru. From the progression in Fig. 3e it is evident that Rh should dilute SOC more efficiently than Ru: the black markers trace the interpolation between the values of Ir and Rh, indicated by the grey line. Meanwhile the modelled impurity potential for Ru ($0.25 \pm 0.05$ eV) prevents successful dilution of SOC. The results in Fig. 3e suggest that the different critical concentrations for the two substituents can be attributed to a common parameter: a value for spin-orbit coupling of $\lambda_{c} \sim 0.42 \pm 0.01$ (indicated as a blue shaded area in Fig. 3e) yields critical concentrations ($x_{\mathrm{Rh}} \sim 0.15$ and $x_{\mathrm{Ru}} \sim 0.3$) that fit well with our experimental observations. Theoretical results presented in Ref. \onlinecite{Watanabe2010} suggest that SOC in Sr$_2$IrO$_4$ is only marginally above the threshold for the insulating state, and that such a small change could drive the transition. The dilution of spin-orbit coupling is therefore found to provide a compelling theoretical picture of the transition.

\begin{figure*}
\centering 
  \includegraphics[width =7in]{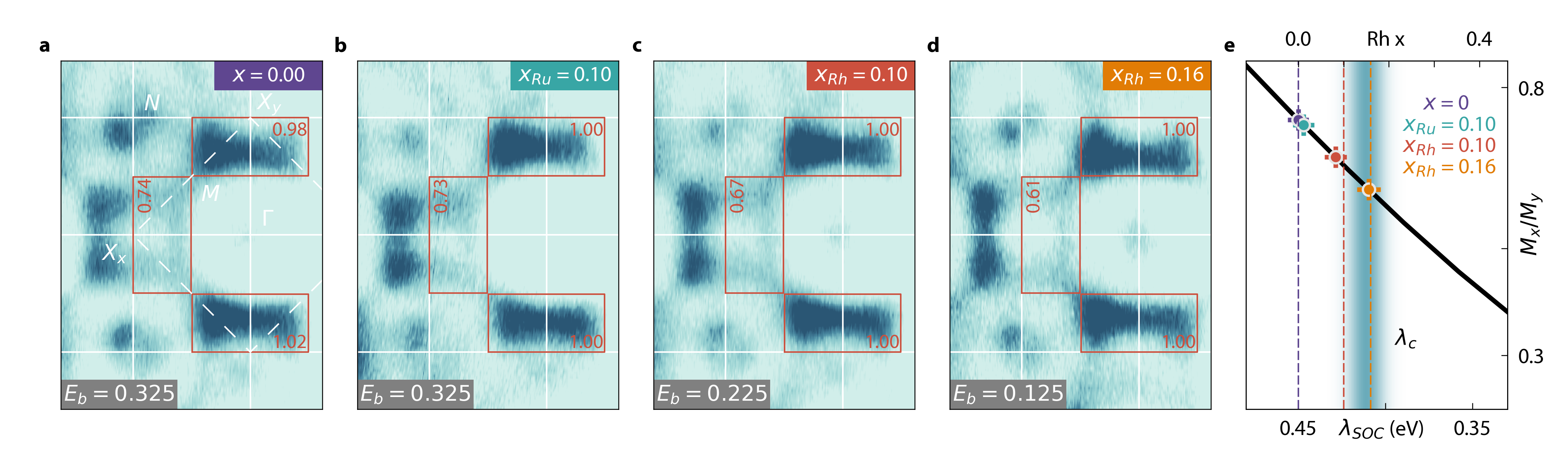}
 \caption{{\bf Observation of the reduction of SOC via the ARPES dipole matrix element.} {\bf a-d} Constant energy maps for different concentrations of $x_{Rh}$, using $\sigma$-polarized light. The constant energy maps are integrated over 150 meV to improve numerical accuracy, and taken at an energy such that the size of the pocket around $X$ is the same for all concentrations. The relevant states used for the analysis are indicated using the red boxes, and their integrated values are shown within. All data are taken at 64 eV, with temperatures at 120 K for $x = 0$ and $x_{\mathrm{Ru}} = 0.1$, 70 K for $x_{\mathrm{Rh}} = 0.1$, and 20 K $x_{\mathrm{Rh}} = 0.16$, all chosen to mitigate the effects of charging. {\bf e} Calculated ratio of matrix elements for a model of Sr$_2$IrO$_4$ (details in the Supplementary Information), plotted as a function of spin-orbit coupling (black curve). The colored markers indicate the ratio of the experimental values shown in panels {\bf a-d}. The error bars are calculated from the standard deviation over the integrated range in energy. The top axis in {\bf e} indicates the substitution required to produce the spin-orbit coupling value on the bottom axis, predicted by the supercell calculations in Fig. 3e. The dashed lines indicate the expected effective spin-orbit coupling from the nominal concentrations of the Rh substituted samples.}
\label{fig4}
\end{figure*}

Having demonstrated this evolution of SOC via substitution and its ability to provide a natural explanation for the transition in Sr$_2$Ir$_{1-x}$T$_x$O$_4$, we aim to substantiate these predictions experimentally. To establish a convenient metric for SOC, we leverage the symmetry constraints of the photoemission matrix element. Dipole selection rules allow transitions from only certain orbitals: since $d_{xz}$ ($d_{yz}, d_{xy}$) is even (odd) in the experimental scattering plane, states composed of this cubic harmonic are only observable with $\pi$- ($\sigma$) -polarization. As SOC mixes these orbitals into linear combinations prescribed by the $j_{\mathrm{eff}}$ construction \cite{Kim2008}, we quantify SOC by comparing the ratio of even/odd states at strategically chosen points in the Brillouin zone where these symmetry-based selection rules are most well defined. In the absence of SOC, the state along $\Gamma-X_x$ (defined in Fig. 4) in Sr$_2$IrO$_4$ would be of pure $d_{xz}$ character: any photoemission from this state using $\sigma$-polarization must be due to the admixture of $d_{yz}$ and $d_{xy}$ introduced by SOC. More quantitatively, of interest here is the value of $M_{x}^\sigma$, the matrix element at the $X_x$ point, which we normalize in our results through division by $M_{y}^\sigma$. A simulation of this quantity based on an \emph{ab-initio} tight binding model for Sr$_2$IrO$_4$ with variable spin-orbit coupling is shown as a black solid line in Fig. 4e. The model takes into account effects of experimental geometry as well as photon energy and polarization; for further details refer to the Supplementary Information. The curve shows a clear decrease of $M_{x}^\sigma/M_{y}^\sigma$ as a function of spin-orbit coupling, demonstrating the possibility for a direct measure of $\lambda_{SOC}$ via ARPES.

Motivated by the supercell calculations, we investigate the progression of $M_{x}^\sigma/M_{y}^\sigma$ experimentally in a series of Rh and Ru substituted samples. In Fig. 4a-d we plot constant-energy contours for each of the concentrations, as recorded with $\sigma$- polarized light. To compare the different samples, we consider constant energy maps at the energy which places the state of interest at $k_x = \left(\frac{3\pi}{4a}, 0\right)$. Integrating and dividing the ARPES intensity within the indicated regions of Fig. 4a-d yields the ratio  $M_{x}^\sigma/M_{y}^\sigma$. We can proceed to make a quantitative connection with an effective spin-orbit coupling strength by plotting the experimental data points alongside the simulated curve in Fig. 4e. The latter has been normalized to the experimental data for pristine Sr$_2$IrO$_4$, allowing for an effective $\lambda_{SOC}$ strength to be extracted for the Rh/Ru substituted samples. This analysis yields $\lambda_{SOC}$ values of $0.443$ ($x_{Ru} = 0.10$), $0.424$ ($x_{Rh} = 0.10$), and $0.408$ eV ($x_{Rh} = 0.16$). A connection to the supercell calculations can be made through these $\lambda_{SOC}$ values: the associated impurity concentrations in Fig. 3e agree remarkably well with the actual experimental values, made explicit in the case of Rh with the top horizontal axis of Fig. 4e. This confirms the premise of our supercell model and the sensitivity to the impurity potential for successful dilution of $\lambda_{SOC}$. In connection to the MIT, the $\lambda_c = 0.42 \pm 0.01$ eV at  $x_{\mathrm{Rh}} = 0.15$ obtained from Fig. 3e is overlain in Fig. 4e. Generally speaking, $\lambda_c$ can also be a function of filling, $U$, bandwidth, and disorder, among others; thus SOC represents but a single axis within a higher dimensional phase space. As filling, distortions, and disorder are expected to expedite rather than suppress the metal-insulator transition in Ru-substituted samples \cite{Yuan2015,Martins2010,Wang2018, Heidarian2004}, the rate in dilution of SOC emerges as the primary responsible for  the dichotomy in $x_c$ observed for Ru and Rh. This indicates the critical role of SOC in the MIT of Sr$_2$Ir$_{1-x}$T$_x$O$_4$ for both Rh and Ru substitution. 

The combination of SOC-sensitive techniques, and the comparison of Ru and Rh substituted samples, place us in a unique position to comment on the role of SOC in the metal-insulator transition of Sr$_2$IrO$_4$, demonstrating for the first time a SOC controlled-collapse of a correlated insulating phase. Through doing so, as an important corollary to these results, our work conclusively establishes Sr$_2$IrO$_4$ as a relativistic Mott insulator. Additionally, we note that the investigation into mixing spin-orbit coupling discussed in SFig. 3 was calculated for a generic two-site Hamiltonian. As such this mechanism pertains to other systems in which this type of physics appears, such as SOC tuning in Ga$_{1-x}$Bi$_x$As \cite{Fleugel2006} and topological insulators \cite{Xu2011, Sato2011,Brahlek2012, Wu2013, Vobornik2014}. Moreover, the sensitivity of these phenomena to an impurity potential has implications for ongoing efforts to enhance SOC effects in graphene and related systems through adatom deposition and other proximity-related techniques \cite{Weeks2011,Hu2012,Avsar2014,Strasser2015,Barker2019}.

\vspace{3mm}
\subsection{Acknowledgements}
We gratefully acknowledge A. Nocera, M. Franz, and G. A. Sawatzky for review of the manuscript and useful discussions. This research was undertaken thanks in part to funding from the Max Planck-UBC-UTokyo Centre for Quantum Materials and the Canada First Research Excellence Fund, Quantum Materials and Future Technologies Program. The work at UBC was supported by the Killam, Alfred P. Sloan, and Natural Sciences and Engineering Research Council of Canada's (NSERC's) Steacie Memorial Fellowships (A.D.), the Alexander von Humboldt Fellowship (A.D.), the Canada Research Chairs Program (A.D.), NSERC, Canada Foundation for Innovation (CFI), and CIFAR Quantum Materials Program. E.R. acknowledges support from the Swiss National Science Foundation (SNSF) grant no. P300P2\_164649. 

\vspace{3mm}
\subsection{Author Contributions}
B.Z. and A.D. conceived the experiment. B.Z., E.R. and M.M. collected the experimental data. N.X., M.S. and J.D.D. provided experimental support. G.C., S.C., K.U., J.B., H.T. and B.J.K. grew the single crystals studied. B.Z. and R.P.D. performed the data analysis. B.Z performed simulations with input from R.P.D., I.S.E. and A.D.. B.Z., R.P.D. and A.D. wrote the manuscript with input from all authors. I.S.E. and A.D. supervised the project. A.D. was responsible for overall project direction, planning, and management.

\vspace{3mm}
\subsection{Competing interests}
The authors declare no competing interests.

\vspace{3mm}
\subsection{Methods}
Single crystals of Sr$_2$Ir$_{1-x}$Rh$_x$O$_4$ were grown with nominal concentrations of $x_{Rh}  = 0.0, 0.10, 0.16, 0.22$ and measured with electron probe microanalysis to be within 0.01 of their nominal concentration. Crystals of  Sr$_2$Ir$_{1-x}$Ru$_x$O$_4$ were grown with nominal concentrations of $x_{Ru} = 0.10, 0.20, 0.40$. Measurements were carried out at the SIS beamline at the Swiss Lightsource (Rh substituted samples) and at the Merlin beamline at the Advanced Lightsource (Rh and Ru substituted samples). All measurements were done on freshly cleaved surfaces, where the pressure during measurement and cleaving was always lower than $3.3\cdot10^{-10}$ mbar. Measurements used for inference of spin-orbit coupling values were performed with 64 eV photons, using light polarized perpendicular to the analyzer slit direction ($\sigma$-polarization). The rotation axis of the manipulator for the acquisition of the Fermi surface was parallel to the slit direction. The sample was mounted such that the Ir-O bonds ($\Gamma-X$) were aligned to this axis of rotation. Temperatures were chosen as low as possible while mitigating the effects of charging and are reported in the figure captions. A tight-binding model was constructed from a Wannier orbital calculation using the Wannier90 package \cite{Mostofi2014}. The Wannier90 calculations were performed on results from density functional theory calculations done with the {\sc Wien2k} package \cite{Blaha2018, Kunes2010}. Supercell and matrix element calculations were performed using the \texttt{chinook} package \cite{Day2019}. Further details can be found in the Supplementary Information. The DOS calculations presented in Fig. 3 were performed with the {\sc Wien2k} package. The supercell configuration assumed a single layer with 8 TM ions per unit cell. The presented results at $x = 0.25$ are similar to those found for $x = 0.125$ and $x = 0.5$. \\\\

\vspace{3mm}
\subsection{Data Availability}
The data represented in Figs. 2 and 3 are available as source data in Supplementary Data 2 and 3. All other data that support the plots within this paper and other findings of this study are available from the corresponding author upon reasonable request.

\end{document}